\documentclass[12pt]{article}
\usepackage{amsmath}
\usepackage{amssymb}
\usepackage{geometry}
\begin{document}
\begin{center}
\begin{Large}
{\bf The Inflaton and Time in the Matter-Gravity System}
\end{Large}
\\[1cm]

{ A.\ Tronconi
 \footnote{tronconi@bo.infn.it},
G.P.\ Vacca\footnote{vacca@bo.infn.it} and 
G.\ Venturi\footnote{armitage@bo.infn.it}}\\[0.5cm]

Dipartimento di Fisica - Universit\`a di Bologna and
INFN - Sezione di Bologna,\\
via Irnerio 46, 40126 Bologna, Italy
\end{center}

\vskip 0.5cm
\begin{abstract}
The emergence of time in the matter-gravity system is addressed
within the context of the inflationary paradigm.
A quantum minisuperspace-homogeneous minimally coupled
inflaton system is studied with suitable initial conditions
leading to inflation and the system is approximately solved in the
limit for large scale factor. Subsequently
normal matter (either non homogeneous inflaton modes or lighter matter)
is introduced as a perturbation and it is seen that its presence requires
the coarse averaging of a gravitational wave
function (which oscillates at trans-Planckian frequencies) having
suitable initial conditions. Such a wave function, which is common for all
types of normal matter, is associated with a ``time density'' in the sense
that its modulus is related to the amount of time spent in a given interval
(or the rate of flow of time). One is then finally led to an effective
evolution equation (Schr\"odinger Schwinger-Tomonaga) for ``normal'' matter.
An analogy with the emergence of a
temperature in statistical mechanics is also pointed out.
\end{abstract}
\section{Introduction}
A fundamental property of the interaction of matter and gravity is its
spacetime reparametrisation invariance. In particular invariance under time
reparametrisation leads to the constraint that the sum of the matter
and gravitational Hamiltonians is zero. This implies that time does not
appear in the Hamiltonian formulation of the matter-gravity
action. However, this does not mean that time does not appear at all in the
classical equations of motion, since they involve time
derivatives of the dynamical parameters, but rather that there is no absolute
time and no clock external to the universe.

Such a feature is maintained and becomes even more evident in the canonical
quantisation of gravity within the superspace approach~\cite{wheeler}.
In such an approach the spacetime dynamical variables are the three geometries
of spacelike surfaces with their conjugate momenta, while
matter is described by the corresponding fields and their conjugate momenta.
As a consequence the classical equations correspond to
geodesic equations in the manifold of all three
geometries~\cite{wheeler, dewitt} (superspace)
modified by the presence of a force term. Canonical
quantisation then leads to a Schr\"odinger-like equation (Wheeler-DeWitt,
hereafter WDW) without any time derivatives and a corresponding wave function
satisfying it.

Clearly, if one considers as a starting point the complete quantum system,
it must be possible to re-obtain the usual classical Einstein
and quantum matter (Schwinger-Tomonaga) equations. This last step, however,
does not appear to be immediate and has raised considerable interest.
In particular the introduction of time has been examined within the
Born-Oppenheimer approximation and the semiclassical approximation to gravity
both without~\cite{banks} and with~\cite{brout,bv} the backreaction of matter.
Further it has also been discussed in quite diverse contexts such as through
an examination of transition matrix elements of radiative processes occurring
in the cosmos~\cite{bp} or through interaction
with the environment~\cite{briggs,ccs} or decoherence~\cite{habib}.

The scope of this paper is to examine the above problem by studying the
matter-gravity quantum system within the context of a
mini-superspace model containing a minimally coupled homogeneous scalar field
(inflaton) which is known to lead to inflation
~\cite{begs,guth,linde,fvv,bellido}
together with normal matter (which could be the non-homogeneous modes
of the inflaton field or generic inflaton decay
products appearing as lighter matter fields).

In the next section we shall study the general minisuperspace-homogeneous
inflaton system and obtain an approximate solution, for large scale factor
(inflation), suitable for some analytic considerations.
Subsequently we shall illustrate  the way the presence of normal
matter leads to the requirement of an evolution (or time) which is
obtained through a coarse graining and relating a gravitational part
of the wave function to a measure of temporal density. Lastly, in
section four, our results are summarized and discussed.
\section{The gravity-matter system}
As we have anticipated in the introduction we consider the
gravity-matter interaction and work with a simple model having an approximate
high degree of symmetry. To be definite we shall consider an inflationary 
cosmology driven by an homogeneous inflaton scalar field, with
the presence of extra matter degrees of freedom, which will be treated as a
small perturbation.
Our goal is to study the effective quantum behavior of these latter degrees
of freedom in the background of the homogeneous inflaton-gravity system.
After having given the classical description of the full
system we shall consider the quantized system described by the corresponding WDW
equation. The study of the consequences of taking an approximate quantum solution
will be of interest for our purposes.
\subsection{The classical system}

We start with a real massive scalar field (inflaton) and other
unspecified matter fields minimally coupled to
gravity which is described by a Robertson-Walker minisuperspace.
The scalar field of mass $m$ is the homogeneous inflaton field $\phi$,
typical of chaotic inflation models, while matter fields
are generally inhomogeneous, and of typical mass $\mu \le m$.

On taking $ds^2=-dt^2+a^2(t)d{\overrightarrow r}^{2}$, the line element in a
flat 3-space, we write the total action with $\hbar=c=1$:
\begin{eqnarray}\label{claction}
S&=&\frac{1}{2}\int dt \left\{ -M^2 a \dot a^2+a^3
\left( \dot \phi^2-m^2\phi^2 \right) + L_\mu \right\}
\end{eqnarray}
where $M$ is the Plank mass and $L_\mu$ is the matter lagrangian,
which depends on the scale factor $a$.
Let us note that the volume factor has been
absorbed by the scaling $a \to a V^{1/3}$. Since $a$ acquires the
dimension of a length, the ${\overrightarrow r}$ variable is dimensionless
just as any momenta ${\overrightarrow k}$ arising from a Fourier analysis,
further the scalar field $\phi$ has the dimension of a mass.

The classical Hamiltonian is:
\begin{eqnarray}
H&=&-\frac{1}{2} \frac{\pi_{a}^2}{M^2 a}+\frac{1}{2 a^3}
\left(\pi_{\phi}^2+m^2 \phi^2 a^6\right)+ H_\mu 
\equiv H_{G}+H_{I}+H_\mu
\label{classicH}
\end{eqnarray}
with $\pi_{a}=-M^2 a \dot a$, $\pi_{\phi}=a^3 \dot \phi$.

\subsection{The quantum system}
The starting point of all our considerations is the quantized matter-gravity
system. We shall consider in this section only the homogeneous inflaton degree of
freedom minimally coupled to gravity, leaving to section $3$ the discussion
of the quantum effects on the other matter fields.
At the quantum level there is no time in the description of the
system since the Hamiltonian, generating the dynamics, is a constraint which
annihilates the physical states. The latter can be written as wave functions
on considering, for example, the $(a,\phi)$ representation.
The dynamics is described by a second order partial differential equation
which can be interpreted as giving a flow in $a$ of some distribution in
$\phi$ given for an initial $a_0$.

Physically we shall be interested in configurations with large $a$, typical 
of a situation which belongs to an advanced stage of the inflationary phase.
At the same time we shall have a constraint on the kind of inflaton field
quantum distribution which must be compatible with such an inflationary regime.

We therefore start from the quantum (WDW) Einstein equation in the
$(a,\phi)$ representation:
\begin{eqnarray}\label{WDW}
&\langle a|\otimes \langle \phi|\left(\hat H_{G}+\hat H_{I}\right)|
\Psi\rangle\equiv&   \nonumber\\
&\equiv\left(\frac{1}{2 M^2}\frac{\partial^2}{\partial a^2}
\frac{1}{a}-\frac{1}{2 a^3}\frac{\partial^2}{\partial \phi^2}+
\frac{m^2 a^3}{2} \phi^2\right)\Psi(a,\phi)=0&
\end{eqnarray}
where, for convenience, we have taken a particular ordering for the
first term ($\frac{\pi_{a}^2}{a}\rightarrow\hat\pi_{a}^2 \frac{1}{\hat a}$)
and $\Psi(a,\phi)$ is the total scale factor-inflaton wave function.
Let us note that different orderings, for example manifestly Hermitian
ones, will differ by terms of the form
$(i/a^2) \hat{\pi}_a$ or $\hat{\pi}_a (i/a^2)$ which, as we shall
later verify, are irrelevant (non-leading) for $a$ large.

It is clear that for a positive definite inflaton Hamiltonian the
wave function $\Psi$ will tend to have a strongly oscillatory
dependence on $a$.
The oscillatory period will be so small that natural scales for the
matter dynamics in $a$ will be much longer and a coarse graining,
which can be, for example, chosen as an averaging over a period of
oscillation in $a$,
will be necessary to study the effective matter dynamics.
A statistical interpretation as given in studies of the probability
distributions in the hydrogen atom \cite{rowe} gives further insight.
We shall develop this idea in section $3$ and concentrate in the following
on studying the oscillatory behavior of the ``universe'' wave function.

It is convenient to write an expansion on a complete basis of the
Hilbert space of the states of the inflaton field
associated with the Hamiltonian $\hat H_{I}$:
\begin{equation}\label{expansion}
\Psi(a,\phi)=a \sum_{n}\psi_{n}(a) u_{n}(a,\phi)\equiv a\:\psi(a)
\sum_{n}c_{n}(a) u_{n}(a,\phi)\equiv a\: \psi(a)\: u(a,\phi);
\end{equation}
on using the eigenstates of the inflaton Hamiltonian $\hat H_{I}$
one has:
\begin{equation}\label{eigenstates}
\hat H_{I}|u_{n}\rangle=m \left(n+\frac{1}{2}\right)|u_{n}\rangle.
\end{equation}
Substitution of (\ref{expansion}) into (\ref{WDW}) leads to
(henceforth $\partial_{a}\equiv \frac{\partial}{\partial a}$):
\begin{eqnarray}\label{PDE}
&u(a,\phi)\:\partial_{a}^2 \psi(a)+ 2\: \partial_{a}u(a,\phi)
\:\partial_{a}\psi(a)+\psi(a)\:\partial_{a}^2 u(a,\phi)+{}&\nonumber\\
&{}+2 M^2 \psi(a)\left(-\frac{1}{2 a^2}
\frac{\partial^2}{\partial \phi^2}+\frac{m^2 a^4}{2}
\phi^2\right) u(a,\phi)=0.&
\end{eqnarray}

Let us comment on the expansion in (\ref{expansion}), and its
statistical interpretation in quantum mechanics.
We note that $|c_n|^2$ is the probability that a state of $n$ quanta is
created (observed) and the $c_n$ arose through the extraction of a common
factor $\psi(a)$ from the $\psi_n(a)$. That is the superposition of the
different $n$ quantum number configurations occurred after extracting the
common factor from the sum of the products of each $n$ quantum state
with its corresponding universe wave function $\psi_n(a)$.
Further we expect the homogeneous inflatonic matter initial state to be such
as to lead to inflation.
Thus we assume that the initial conditions are associated
with the creation (considered as the inverse of a decay
process) of a large average number $\bar n$ of quanta in an initial
interval of $a$. 
This can then be imagined as occurring  through a
Poisson process so that the initial state associated with the interval
of $a$ is related to a Poisson distribution.

Thus, instead of having a sequence of random creations of $n$ quanta
(with average $\bar n$) corresponding to a sequence of inflationary
processes in successive infinitesimal intervals of $a$ ,
we consider a statistical average, that is a superposition of the different
possible quanta numbers which can occur in the interval of $a$ leading to
inflation. Subsequently $a$ will undergo a rapid change due to inflation.

This distribution is familiar in quantum mechanical oscillators.
It is  associated with coherent states which, for large average occupation
numbers, describe an almost classical behavior.
We shall discuss the above assumption in the next section and
find that for large $a$ it is a reasonable one, independently of any
possible interpretation of its origin.
\subsection{Approximating the quantum inflaton-gravity system}
This section is devoted to the construction of an approximate solution of
the WDW which corresponds to an inflationary situation.
As a next step, in the following section, we shall make use of the
results of this section in order to obtain some analytical insights on the
dynamics of other matter degrees of freedom and show how a concept of time arises
naturally, a phenomenon which in any case appears to be independent of the 
approximation employed.

To extract an approximate solution two procedures are possible:
either try to solve directly the partial differential equation
(\ref{PDE}) or make an ansatz for $u(a,\phi)$ and show that it
leads to an approximate solvable differential equation for $\psi(a)$.
We shall choose the latter approach.

For convenience we rewrite the inflaton Hamiltonian
$\hat H_{I}(a)\equiv \langle a|\hat H_I|a \rangle$
in terms of inflaton creation and annihilation operators:
\begin{equation}\label{Hi}
\hat H_{I}(a)\equiv\frac{\hat\pi_{\phi}^2}{2 a^3}+
\frac{m^2 a^3}{2}\hat\phi^2=m \left( b^{\dag} b+\frac{1}{2}\right)\, ,
\end{equation}
where
\begin{eqnarray}\label{bbstar}
b=\sqrt{\frac{m a^3}{2}}\left(\hat\phi+\frac{i}{m a^3}
\hat\pi_{\phi}\right) \, , \quad 
b^{\dag}=\sqrt{\frac{m a^3}{2}}\left(\hat\phi-
\frac{i}{m a^3}\hat\pi_{\phi}\right)
\end{eqnarray}
and the ansatz for $u(a,\phi)$ will be:
\begin{equation}\label{ansatz}
u(a,\phi)=\langle \phi| \alpha (a)\rangle\, ,
\end{equation}
where $|\alpha(a)\rangle$ is a coherent state of the inflaton, defined by
\begin{equation}\label{defcoherent}
b|\alpha (a))\rangle=\alpha(a) |\alpha(a)\rangle\, ,
\end{equation}
with $\alpha(a)=\sqrt{\frac{m \phi_{0}^{2}a^3}{2}}$.
From eqs. (\ref{ansatz}) and (\ref{defcoherent})
$u(a,\phi)$ must satisfy the differential equation:
\begin{equation}\label{eqcoherent}
\left(\frac{1}{m a^3}\frac{\partial}{\partial\phi}+
\phi-\phi_{0}\right) u(a,\phi)=0\, ,
\end{equation}
where $\phi_{0}$ is a free parameter. Equation
(\ref{eqcoherent}) has the familiar normalized solution:
\begin{equation}\label{u}
u(a,\phi)=\left(\frac{m a^3}{\pi}\right)^{\frac{1}{4}}
\exp\left[-\frac{m a^3}{2}(\phi-\phi_{0})^2\right]\, ,
\end{equation}
which is a simple gaussian peaked around $\phi_{0}$ with a width which
decreases as the scale factor $a$ increases.
The values of the parameter $\phi_0$ are constrained in order to
lead to an inflationary regime \cite{fvv} ($\bar n = \alpha(a)^2$), since
such a quantum coherent state description can be related
to the classical analysis of the inflaton system.

Let us now study the consequences of such a choice for $u$.
If we substitute the expression (\ref{u})
into equation (\ref{PDE}) and calculate the contributions of
the different derivatives we obtain:
\begin{eqnarray}
&&\partial_{a} u(a,\phi)=\left[\frac{3}{4 a}-\frac{3 m a ^2}{2}
(\phi-\phi_{0})^2\right]u(a,\phi)\label{est1}\, ,\\
&&\partial_{a}^2 \:u(a,\phi)=\left[-\frac{3}{16 a^2}-\frac{21 m a}{4}
(\phi-\phi_{0})^2+\frac{9 m^2 a^4}{4}(\phi-\phi_{0})^4
\right]u(a,\phi)\label{est2}\, ,\\
&&\frac{\partial^2}{\partial\phi^2}u(a,\phi)=
\left[-m a^3+m^2 a^6
(\phi-\phi_{0})^2\right]u(a,\phi)\label{est3}\, .
\end{eqnarray}
A key point is that we are interested in solutions of (\ref{WDW}),
describing an advanced stage of inflation ($a\gg 1 $),
since in this phase we wish to study how normal matter behaves.
In this limit all the terms in the
expressions (\ref{est1}-\ref{est3}) which are of the form
$(\phi-\phi_{0})^n u(a,\phi)$ with $n>0$,
are non-leading by powers of $a$ with respect to $u(a,\phi)$ since
\begin{equation}\label{dis}
\max \left[|\phi-\phi_{0}|^n u(a,\phi)\right]\sim
\left(\frac{n}{m a^3}\right)^{\frac{n}{2}}
\exp\left[-\frac{n}{2}\right]\max \left[u(a,\phi)\right]\, ;
\end{equation}
in other words one has:
\begin{eqnarray}
&&\partial_{a} u(a,\phi)=O( a^{-1}) \; u(a,\phi)\, \label{u1eq},\\
&&\partial_{a}^2 \:u(a,\phi)=O(a^{-2}) \; u(a,\phi)\, \label{u2eq}, \\
&&\frac{\partial^2}{\partial\phi^2}u(a,\phi)=O(a^{3}) \; u(a,\phi) \;.
\end{eqnarray}
On just retaining the leading contributions in (\ref{PDE}) one has:
\begin{equation}\label{leadPDE}
\left[\partial_{a}^2 \psi(a)+m^2 M^2 \phi^2
\:a^4 \psi(a)\right]\: u(a,\phi)=0
\end{equation}
where $u(a,\phi)$ has support in a tiny region around $\phi_{0}$, due to the
large values of $a$. We have also used the fact that any term of
$O(1/a) \psi(a)$ is negligible with respect to $\partial_a \psi(a)$
for large $a$, as can be verified on using the solution found for $\psi(a)$.
Further additional terms such as $\partial_a (1/a^2)$ or
$(1/a^2) \, \partial_a$, due to different orderings in the
gravitational kinetic term, are non-leading when acting on
$a \ \!\! u(a,\phi)$ (see eq. (\ref{u1eq}) and (\ref{u2eq}) )
and are also negligible with respect to  $\partial_a^2
\psi$ when acting on the solution found for $\psi(a)$. 
Therefore the choice of ordering is irrelevant for large $a$. 
Finally one may rewrite (\ref{leadPDE}) as:
\begin{equation}\label{qairy}
\partial_{a}^2 \psi(a)+m^2 M^2 \phi_{0}^2\: a^4 \psi(a)=0
\end{equation}
or on changing to the variable $y=a^2$ (again apart from a non-leading
term for $a$ large),
\begin{equation}\label{AIRY}
\frac{\partial^2}{\partial y^2}\psi(y)+
\frac{m^2 M^2 \phi_{0}^2}{4}\:y\: \psi(y)=0 \, .
\end{equation}
A general solution, in terms of the Airy functions $A_i$ and $B_i$,
for eq. (\ref{AIRY}) is:
\begin{equation}\label{genAIRY}
\psi(y)=C_{1} Ai(y)+C_{2}Bi(y)\rightarrow y^{-\frac{1}{4}}
\left[D_{1}\sin \frac{m M \phi_{0}}{3}
y^{\frac{3}{2}}+D_{2}\cos \frac{m M \phi_{0}}{3}y^{\frac{3}{2}}\right]
\end{equation}
as $y\to\infty$, or:
\begin{equation}\label{genAIRYa}
\psi(a)\sim  a^{-\frac{1}{2}}\left[D_{1}\sin \frac{m M \phi_{0}}{3}
a^{3}+D_{2}\cos \frac{m M \phi_{0}}{3}a^{3}\right]
\end{equation}
for $a\gg1$ where $D_1$ and $D_2$ are complex numbers.
The oscillatory behavior is encoded in $\psi$, even if at an
approximate level (the solution is not exact).
Let us note that the $D_i$ ($C_i$) in (\ref{genAIRY}) can be determined by
the initial conditions. For example, $C_1=-i$ and $C_2=1$ corresponds to
the Vilenkin initial wave function of the universe, while if
$C_1=1$ and $C_2=0$ one has the Hartle-Hawking choice.

One may ask what would be the effect of a perturbation such as the
addition of a small contribution of $n$ inflaton quanta. Because of the rapid
oscillatory behavior of the gravitational wave function and the fast increase
of $\bar n$ as $a \to \infty$ such a contribution will be quickly washed out.
Thus our ``coherent'' state will remain dominant during inflation.
Nonetheless some correction are certainly expected.
For example in the classical limit (one has the classical Einstein equation)
there are nonleading non power-like corrections to the oscillatory phase which
are expcted to be of a logarithmic nature.
We shall comment more on this in the following section.  
\section{The appearance of time in the description of matter}
Let us at this point discuss the dynamics of ``normal'' matter fields.
The quanta of these fields, associated with matter particles,
are assumed to contribute as a small perturbation
in the total Hamiltonian where the inflaton (homogeneous mode condensate)
and gravity are dominant and whose classical expression was given in
(\ref{classicH}). 
In order to study this quantum problem one has to enlarge the corresponding
Hilbert space of the physical states, in which the inflaton-gravity
quantum state is only negligibly affected by the presence of ``normal''
matter, and consider the following extended quantum equation for the
factorized wave function:
\begin{eqnarray}\label{WDWmat}
&\left[\frac{1}{2 M^2}\frac{\partial^2}{\partial a^2}
\frac{1}{a}-\frac{1}{2 a^3}\frac{\partial^2}{\partial \phi^2}+
\frac{m^2 a^3}{2} \phi^2 + \hat{H}_\mu \right]\Psi \, \chi
\equiv&\nonumber\\
&\equiv\langle a|\otimes\langle \phi|\otimes\langle \varphi|
\left(\hat H_{G}+\hat H_{I}+\hat H_{\mu}\right)
|\Psi\rangle\otimes|\chi\rangle=0& \, ,
\end{eqnarray}
where we have collectively denoted with $\varphi$ all the matter fields
(bosonic and fermionic) and $\chi(\varphi,a)$ describes their
quantum state in some representation.

The first term in eq. (\ref{WDWmat}) $\hat H_{G}$ acts on both
$\Psi(a,\phi)$ and $\chi(\varphi,a)$,
the last ($\hat H_{\mu}$) only on $\chi$ and $\hat H_{I}$
only on $\Psi$.
On using eq. (\ref{WDW}) for $\Psi$ we can obtain an equation for the
``normal'' matter wave function $\chi$:
\begin{equation}
2 \:\partial_{a}\left(\frac{\Psi}{a}\right) \:\partial_{a}\chi+
\frac{\Psi}{a} \partial_{a}^{2}\chi+2 M^2\Psi\:
\hat H_{\mu}(\varphi,a)\chi=0
\end{equation}
where, for sake of compactness,
$\hat H_{\mu}(\varphi,a)\equiv 
\langle a|\otimes\langle\varphi| \hat H_{\mu}|a
\rangle\otimes|\varphi\rangle$. Further on defining
$\tilde \psi(a,\phi)\equiv \psi(a) u(a,\phi)$ and bearing in mind
(\ref{expansion}) one has:
\begin{equation}\label{eqmat1}
2\frac{\partial_{a}\tilde\psi}{\tilde\psi}\:\partial_{a}\chi+
2 a M^2\: \hat H_{\mu}(\varphi,a)\, \chi+
\partial_{a}^2\chi=0.
\end{equation}
Let us now examine the different terms in (\ref{eqmat1}) while assuming the
presence of a finite number $N$ of quanta of normal matter.
It is convenient to consider a free matter (bosonic or fermionic) field.
Then it is clear that the eigenvalues of the Hamiltonian will be independent
of $a$ as $a \to \infty$. For example a state on $N$ bosonic quanta of momentum $k$
leads to a contribution $\sim \omega_{k} \left(N+\frac{1}{2}\right)$
where $\omega_{k}^2=\frac{k^2}{a^2}+\mu^2$
(apart from a normal ordering subtraction).
The last term is then proportional to:
\begin{equation}
\left(\sum_{L}\langle N|\overleftarrow\partial_{a}^2|L
\rangle\langle L|\overrightarrow\partial_{a}^2|N\rangle\right)^{\frac{1}{2}}
\propto \frac{N^2}{a^2}
\end{equation}
and is negligible with respect to the second for large $a$.

The first term in (\ref{eqmat1}) contains all the information related to
the behavior of the gravity-inflaton wave function which is highly
oscillatory in $a$ with a typical period much less than a Planck time (length)
(see below).
Since any physical phenomena of interest for matter are related to scales
above (or at least not below) the Planck time $M^{-1}$,
it is very natural to obtain an effective equation by means
of a course graining. In particular we shall implement it by averaging
over an oscillation period in $a$.
\subsection{Approximate analysis}
In order to compare the first two terms of (\ref{eqmat1})
we consider the explicit expression of the approximate solution for
$\Psi=a \, \tilde{\psi}$ derived in the preceding section and write:
\begin{equation}\label{dpsi}
\frac{\partial_{a}\tilde\psi}{\tilde\psi}\equiv
\frac{\partial_{a}\psi(a)}{\psi(a)}+\frac{\partial_{a}u(a,\phi)}{u(a,\phi)}.
\end{equation}
On recalling eq. (\ref{est1}) the second term behaves as
$\frac{\partial_{a}u(a,\phi)}{u(a,\phi)}\sim \frac{1}{a}$
while the first term is asymptotically given by
\begin{equation}\label{dpsi/psi}
 \frac{\partial_{a}\psi(a)}{\psi(a)}=\frac{\partial y}{\partial a}
\frac{\partial_{y}\psi(y)}{\psi(y)}\sim
\frac{m M \phi_{0}\, a^{\frac{3}{2}}}{\psi(a)}\left[D_{1}\cos \frac{m M \phi_{0}}{3}
a^{3}-D_{2}\sin \frac{m M \phi_{0}}{3}a^{3}\right].
\end{equation}
The crucial point is now to estimate the ratio (\ref{dpsi/psi}):
$\psi(a)$ and its derivative are products of a highly 
oscillating function of $a$  with some power of $a$.
Let us consider the $a$-dependent period of oscillation $\Delta a$
\begin{equation}\label{deltaa}
\Delta a\sim \frac{2 \pi}{m M \phi_{0} a^{2}}.
\end{equation}
Since $M$ is the Plank mass,
$m \sim 10^{-6}M \ll M$, $\phi_{0}\sim M$ and $a \gg M^{-1}$
(by many orders of magnitude) during the inflationary phase for $a$ large, one obtains
the inequality
\begin{equation}\label{stimadeltaa}
\Delta a \ll M^{-1}.
\end{equation}
Thus one may estimate (\ref{dpsi/psi}) by averaging over an interval
$[a, a+\Delta a]$ which is much shorter than a Planck length:
\begin{equation}\label{cgint}
I_{av}=\frac{1}{\Delta a}\int_{a}^{a+\Delta a} d \alpha\:
(m M \phi_{0}\, \alpha^2) \frac{\left[D_{1}\cos \frac{m M \phi_{0}}{3
}\alpha^{3}-D_{2}\sin \frac{m M \phi_{0}}{3}\alpha^{3}\right]}
{\left[D_{1}\sin \frac{m M \phi_{0}}{3
}\alpha^{3}+D_{2}\cos \frac{m M \phi_{0}}{3}\alpha^{3}\right]}.
\end{equation}
The coarse-graining integral (\ref{cgint}) is well defined for any complex
$D_1$ and $D_2$, except for a zero measure case where their ratio is real,
since the denominator in the integrand may vanish. To obtain a finite result 
one may then consider the principal value of the integral.
We shall see that, except for the latter case and  depending on the initial 
conditions discussed at the end
of section $2$, one is able to introduce a
time evolution. In particular one finds that 
Vilenkin type wave functions allow one to introduce a time, while
Hartle-Hawking ones do not.
On defining a new integration variable
$x=\frac{m M \phi_{0}}{3}\alpha^{3}$ one may write:
\begin{align}\label{cgintx}
I_{av}
&=\frac{m M \phi_0}{2 \pi} a^2 \int_{0}^{2 \pi} dx \;
\frac{\left[D_{1}\cos x-D_{2}\sin x\right]}{\left[D_{1}\sin x
+D_{2}\cos x\right]}\, ,
\end{align}
which can be evaluated in the complex plane by using Jordan's Lemma and
the Residue Theorem. In fact on setting $z=\exp{i x}$ one has 
$\cos x=\frac{1}{2}\left(z+\frac{1}{z}
\right)$, $\sin x=\frac{1}{2 i}\left(z-\frac{1}{z}\right)$,
$dx=-\frac{i}{z}dz$ and the integral (\ref{cgintx}) becomes:
\begin{equation}\label{cgintz}
I_{av}=\frac{m M \phi_0}{2 \pi i} a^2\oint \frac{dz}{z} \,
\frac{i D_1 z^2+i D_1-D_2 z^2+D_2}{D_1 z^2-D_1+i D_2 z^2+i D_2} 
\end{equation}
The function in the integrand (let us denote it by $F(z)$) has three
singularities, at $z_0=0$ and
$z_{1,2}=\pm \sqrt{\frac{D_1-i D_2}{D_1+i D_2}}\equiv
\pm A e^{i\theta}$: $z_0$
will always lie within the path of integration
(the circumference of radius 1 centred on the origin of the complex plane)
while $z_{1,2}$ generally will not.
Only if $D_1$ and $D_2$ have a real ratio one has $A=1$ and
$z_{1,2}$ are on the circumference.
If one evaluates the residues of $F(z)$ at $z_{0,1,2}$ one obtains:
\begin{align}\label{res}
Res F(z)|_{z=z_0}=-i \, , \quad 
Res F(z)|_{z=z_{1,2}}=i
\end{align}
To summarize, one can distinguish between 3 cases:\\
when $0<A<1$:
\begin{equation}\label{Amin1}
I_{av}=m M \phi_0\, a^2 
\left(Res F(z)|_{z=z_0}+Res F(z)|_{z=z_1}+Res F(z)|_{z=z_2}\right)=
i \, m M \phi_0\, a^2
\end{equation}
when $A>1$:
\begin{equation}\label{Amag1}
I=m M \phi_0\, a^2 \left(Res F(z)|_{z=z_0}\right)=
-i \, m M \phi_0\, a^2
\end{equation}
when $A=1$ the calculations are slightly different since one has to deform
the path of integration to avoid the two singularities and
then let the deformation tend to zero. For instance,
the path could turn around $z_{1,2}$ following:
$z_{1,2}(\gamma)=z_{1,2}+\epsilon e^{i\gamma}$
with $\gamma$ going from $\theta-\frac{\pi}{2}$ to
$\theta-\frac{3 \pi}{2}$ around $z_1$ 
and from $\theta+\frac{\pi}{2}$ to
$\theta-\frac{\pi}{2}$ around $z_2$:
\begin{align}\label{intaug1}
I&=m M \phi_0\, a^2 \left(Res F(z)|_{z=z_0}\right)-
\lim_{\epsilon\rightarrow 0}\left[ \frac{m M \phi_0}{2 \pi i
}a^2\left( \int_{z_1(\gamma)}dz F(z)+\int_{z_2(\gamma)}dz F(z)\right)
\right]=\nonumber\\
&=-i\, m M \phi_0\, a^2-\left[ \frac{m M \phi_0}{2 \pi i
}a^2\left(\pi+\pi\right)\right]=0.
\end{align}
Thus,unless $A=1$, coarse-graining of the ratio (\ref{dpsi/psi})
leads to a non zero result, in which case the dominant contribution 
in (\ref{dpsi})
comes from the first term and one has:
\begin{equation}
\frac{\partial_{a}\tilde\psi}{\tilde\psi}\sim\pm i\,
m M \phi_{0}\, a^{2}.
\end{equation}

We must further estimate $\partial_{a}\chi$, again for N quanta,
obtaining for its magnitude:
\begin{equation}
\left[\langle N|\overleftarrow\partial_{a}\overrightarrow
\partial_{a}|N\rangle\right]^{\frac{1}{2}}\equiv
\left[\sum_{L}\langle N|\overleftarrow\partial_{a}|L
\rangle\langle L|\overrightarrow\partial_{a}|N\rangle
\right]^{1/2}\propto\frac{N}{a}\; .
\end{equation}
As for the last term, we have shown it to be highly suppressed.

On just retaining the two leading terms in eq.(\ref{eqmat1}) one obtains
\begin{equation}\label{emtime}
\pm i \frac{m\phi_{0}}{M}a\:\partial_{a}\chi=
\hat H_{\mu}(\varphi,a)\chi \, .
\end{equation}
Therefore we are left with a parabolic PDE which describes the evolution
of ``normal'' matter with respected to $a$.
On just defining a time $t$ according to 
$\pm i\, m\phi_{0}/M \partial_{log a}\chi \equiv i\, \partial_t \chi$ one
obtains a Schr\"odinger equation describing the quantum
mechanical evolution of matter, or a Schwinger-Tomonaga equation
\begin{equation}\label{schr}
i \, \partial_{t}\chi=
\hat H_{\mu}(\varphi,a)\chi \, .
\end{equation}
Let us note that this is equivalent to interpreting the coefficient of
$i\,\partial_{a}\chi$ on the l.h.s. of (\ref{emtime}) as
$\dot a$, the rate of expansion seen by the ``normal'' matter
(observers + clocks).
In particular the approximate analysis employed here shows a DeSitter type
expansion.
In general an observer made of ``normal'' matter, using clocks made of normal
matter, can only measure any change of state with respect to the evolution
of the $t$ introduced above as the parameter of the Schr\"odinger equation
(\ref{emtime}).
In other words it is the presence of normal matter which leads naturally to
the introduction of a time.  
The sign in eq. (\ref{emtime}) depends on the initial conditions
which characterize the solution $\psi(a)$.
With generic conditions, 
one of the two cases is realized and this, in our approach,
is related to the definition of the flow of time.

\subsection{Beyond the approximation}
The result we have obtained is reasonable, especially considering the
approximations made during the inflationary phase.
A different scale dependent evolution seen by matter, such as the one obtained in the
classical limit of the chaotic inflation model,
could be easily recovered by a very mild correction
(logarithmic) to the coefficient of $i\,\partial_a \chi$,
which allowes for the difference between a constant and an approximate
linearly decreasing (in time) Hubble parameter.

For the general case, with a strongly peaked inflaton wave function,
independently of any particular approximation,
the oscillatory wave function in the large $a$ region can be
written as $\tilde \psi(a) =\rho(a)
\exp{ i \beta(a)}$, where $\rho$ is an oscillating real function.
For most initial conditions, as in the approximate solution,
such a function never vanishes except possibly at an extreme of the oscillation. 
Only a zero measure set of initial conditions
will not satisfy this property leading to the same
consequences as already discussed in the previous subsection.
After the course-graining integration one can therefore make the
correspondence:
\begin{equation}\label{general}
\frac{1}{a\, M^2}
\frac{\partial_a \tilde{\psi}}{\tilde{\psi}}\partial_a=
\frac{1}{a\, M^2} \left( \frac{\rho'(a)}{\rho(a)}+
i \beta'(a) \right)\partial_a \approx
\frac{i}{a M^2} \beta'(a) \partial_a \to i\, \partial_t
\end{equation}
where a prime denotes differentiation with respect to $a$,
since after course-graining the term $\rho'(a)/(a\, \rho(a)) \partial_a$
is always negligible and the gradient of the phase is unaffected. From 
this formula one can recover the previous approximate result. Indeed,
for example, for the case of the Vilenkin type wave function
one has $\beta(a)=m M \phi_0 \, a^3/3$.
Again time is defined in terms of the inverse of the gradient of the phase
of the $\psi$ wave function with respect to the scale factor, and in
particular $dt \equiv (a M^2/ \beta'(a))\, da$.

Until now we have worked with the amplitude and not the probability density,
however for large $a$ a probability density analysis leads to the same
result. One may easily convince oneself of this on examining the probability
current, which satisfies the relation
\begin{equation}
J_a\propto \tilde \psi^* \partial_a \tilde \psi-
\tilde \psi \partial_a \tilde \psi^* \approx 
2 i \rho^2 \partial_a \beta \propto \partial_a \tilde \psi.
\end{equation}
since, in the limit we consider and on repeating the previous steps,
the terms leading to differences are negligible.
This means that the density flux is proportional to the gradient of
the phase and the fundamental condition for time to arise, in our
framework, is then a non negligible coarse-grained probability current.

The presence of an effective time evolution for matter arises from a mechanism
similar to one already observed in the analysis of the classical limit
of quantum systems, such as the  hydrogen atom \cite{rowe},
in the sense that the quantum probability as a function of $a$ is similar
to the measure of temporal density in a classical orbit.
This fact has been studied for the stationary quantum eigenstates of the
hydrogen atom (with two particular fixed values of the angular momenta and large principal
quantum number $n$) one of which presents a radial highly oscillatory behavior.
On course graining (in particular on applying the Riemann-Lebesgue Lemma)
one is able to recover the classical trajectory related to the given
angular momenta. Indeed the classical trajectory is related to a classical 
spatial probability distribution of a particle in terms of the inverse
of its speed (the fraction of time spent in a spatial interval is a measure of the
probability density).
There is a deep connection between the above example and the situation
present in the matter-gravity system.

As a side remark we note that for the particular case wherein one performs a
WKB analysis in order to obtain the classical limit of equation (\ref{WDW})
(since this leads to the Hamilton-Jacobi equation),
$\tilde \psi$ is characterized by $\rho \sim (\beta')^{-1/2}$
(with $\beta$ the classical action) which does not oscillate and the
usual relation $\dot a \sim \pi_a/a \sim \beta'(a)/a$ is found.
In any case we stress that the relation in (\ref{general}) is quite general,
a much wider class of states for the inflaton-minisuperspace system can
lead to an effective time evolution of ``normal'' matter.

It is worth noting that we always obtain in the course-grained
differential equation for ``normal'' matter, which constitutes a
perturbation in the whole Hamiltonian system, a factorized term of
the form $f(a) \partial_a \chi$.
In such a term $f(a)$ is related to the variation of the universe wave function
(minisuperspace+inflaton) regarded as a system separate from ``normal'' matter.
It therefore leads to a universal time for all the ``normal'' matter since it
is a common factor for all types of "normal" matter.
One has an analogy with the introduction of a temperature in statistical
mechanics if one thinks of the minisuperspace-inflaton system in dynamical
equilibrium as a sort of ``heat'' bath and normal matter (a perturbation) 
in ``thermal'' equilibrium with the ``bath'' (reservoir), in the sense that
is acquires from the ``bath'' a time (in analogy with temperature in
statistical mechanics).   

A short comment also on a possible highly quantum mechanical situation
wherein the initial condition leads to a coherent superposition of
states described by a wave function of the inflaton $\phi$ consisting
of well separated, almost gaussian, peaks.
A similar definition of time can be extracted on repeating the analysis
performed above using a probability density and its flux,
constructing a density matrix and subsequently tracing over the possible
inflaton states. Indeed the different superimposed states are almost orthogonal
and one is finally left with an average gradient of the phase which
plays an analogous role to the single peak case in the construction of
an effective Schr\"odinger equation for matter. 
The application of course-graining techniques in the
context of quantum cosmology has also been considered by other authors
\cite{ccs,habib}, for example in the study of the wave
function of a minisuperspace model by using the Wigner function approach,
where diverse kind of coarse-graining are discussed in the search for the
emergence of a classical behavior for gravity.

In our quantum mechanical framework it is tempting to say that time
``exists'' only as far as ``normal'' matter or inhomogeneous modes for the
inflaton or metric fluctuations are concerned.
The homogeneous minisuperspace gravity and inflaton condensate system
may still be in a quantum state, which is practically unobservable in
the sense specified above for the other physical degrees of
freedom (``normal'' matter and inhomogeneous modes of inflaton and gravity).
Indeed any observer, made of ``normal'' matter will only see a classical
time dependent scale factor of the universe. Lastly let us emphasize that
the presence of inflatonic  matter driving the universe
is crucial: in a chaotic inflation model for which $\phi_0=0$ \cite{fvv}
one would not obtain suitable conditions for time to evolve.
\section{Conclusions}
As we mentioned in the introduction the problem of the emergence of time
in the context of the matter-gravity system has already drawn
considerable attention.
Nonetheless we feel that the issue is worth revisiting in a cosmological
minisuperspace context, in particular within the inflationary
paradigm \cite{begs,guth,linde,bellido}.
Inflation gives a framework in which to pose basic cosmological questions.

Associated  with inflation is the early universe dominance of a vacuum
energy density effectively described by a hypothetical scalar field
called inflaton.
In particular we have considered a simple model of chaotic inflation
consisting of a minimally coupled massive scalar field in a flat 3-space
with inflation being driven by the homogeneous mode.
We then start from the quantum minisuperspace-homogeneous scalar field
system. Such a quantum system has been previously studied within a
Born-Oppenheimer approach and shown to lead to inflation even for non
classical initial states \cite{fvv}.
Our present approach and goal are quite different: we study this system, in an
inflationary regime, in order to understand its influence on the dynamics of
other degrees of freedom which we call ``normal'' matter. 

The quantum minisuperspace homogeneous-inflaton system cannot be
solved exactly, however some of its properties can be obtained through the use
of a suitable ansatz for the initial
state which allowes us to find an approximate solution for $a$ large.
The results we obtain are not a consequence of our approximation, which
certainly aids us in suggesting and understanding the basic mechanisms, but reflect
a general structure for the effective dynamics of ``normal'' matter. 

Our approximate approach consists in expanding the total matter-gravity
wave function on a basis of the states associated with
different numbers of homogeneous inflaton quanta.
Subsequently, a common minisuperspace wave function is extracted from
the expansion  and the modulus squared of the coefficients
multiplying the states with different number of inflaton quanta is then
interpreted as the probability for the creation of such a state in the
initial interval leading to inflation, bearing in mind, of course, that
such a number is large in order for suitable inflation to take place.

The search for a suitable ansatz for the initial state may be related to
a creation process for a large number of quanta. Such a mechanism can be
considered as analogous to the inverse of a decay process, that is a Poisson
process, without concerning oneself as what it is (the equivalent of the decay
products)that leads to the creation.
The initial conditions are then associated
with a Poisson distribution of states
with diverse numbers $n$ of inflaton quanta and with average number of quanta $\bar n$. 
This can be seen either as an initial
statistical distribution in which each n quantum creation is associated with an
inflationary gravitational wave function or a sequence of inflationary
processes in successive intervals of $a$ which are however so close that $a$
is approximately constant.
In all cases a common part of the gravitational wave function is extracted leaving
the different quanta numbers states with their Poisson weights which
then combine into a coherent state.

The quantity which is then studied is the common gravitational wave function.
It is found to satisfy the same Airy equation for $a$ large,
independently of any ordering chosen for the gravitational kinetic
term, leading to a strong oscillatory behavior having a
frequency many orders of magnitude greater than the Planck one.
The oscillatory behavior of the complete wave function is the main
feature we use and we stress that is not related to the approximation
made but just to initial conditions such as those leading to inflation.

One may at this point introduce normal matter which should be
regarded as a small perturbation with respect to the contribution of the homogeneous
inflaton mode. Again in this context $a$ is considered large and an
effective ``flow'', due to the nature of the gravity-inflaton wave
function, can be associated to the existence of time for normal matter.
In order to obtain this, a ``coarse graining'' was performed so as to
smooth out the effects of the gravitational wave function oscillations at
ultra-Planckian frequencies (for the atomic case see \cite{rowe}),
to which normal matter is insensitive.
It is also worth noting that the common gravitational wave function term
leads to the introduction of a ``time density''
(the magnitude of the gravitational wave function in an interval of $a$
being related to the ``time'' spent in that interval) which is universal,
i.e. independent of the type of normal matter associated with it.
The paradigm of time developed here arises from a novel point of
view, but nonetheless leads to the usual time that one considers in the
WKB limit of the gravity-inflaton system.

Moreover we observe that a new interesting possibility is the emergence
of a time even with gravity and the inflaton in quantum regimes:
it is a consequence of the fact that normal matter cannot see quantum fluctuations below
the Planck size but just experiences an evolution with respect to a
function of the scale factor, associated with the speed of inflation.
In such a framework time only exists for 
normal matter which evolves according to its position on the gravitational wave
function. 

In a sense, and particularly on considering the approximate analysis,
this emergence of a universal time reminds us
of the derivation of a Boltzmann distribution and a
temperature for systems obtained by placing them in contact with a
large heat bath where it is the density of energy levels of the heat bath that is
related to the (common) temperature of systems in contact with it, different energy
level densities for the reservoir being associated with different temperatures.
Analogously, for us, normal matter is in contact with a gravitational wave
function generated by the homogeneous mode inflaton background
(reservoir-``time bath'') and it is that wave function with suitable initial
conditions that generates a ``time density'' and ``flow'' which leads to the
usual evolution of matter corresponding to the value of the matter wave
function for different values of $a$ in the gravitational wave function.

Lastly let us note that we have studied the introduction of time
during the inflationary era with gravitation being driven by the
homogeneous inflaton Hamiltonian. Clearly one may ask what happens at
the end of inflation when presumably all (or most or some) of the
homogeneous modes have decayed into lighter matter.
In such a case gravitation will just be driven by the mean energy
of all matter (including a residual cosmological constant) and
the resulting gravitational wave function will introduce a common
time for any small sub-system which may be regarded as a
perturbation of the whole.
\section*{Acknowledgement}
We wish to thank R. Brout for several clarifying discussions.

\end{document}